\documentclass[prb,nobibnotes,showpacs,amsmath,amssymb]{revtex4}

\usepackage{graphicx}   

\newcommand{\rd}{{\rm d}}

\newcommand{\kB}{k_{\rm B}}
\newcommand{\TC}{T_{\rm C}}

\begin{document}

\title{Factorising numbers with a Bose--Einstein condensate}

\author{Christoph Weiss, Steffen Page, Martin Holthaus}

\affiliation{Institut f\"ur Physik, Carl von Ossietzky Universit\"at, 
              \\ D-26111 Oldenburg, Germany}

\date{March 5,2004}

\begin{abstract}
The problem to express a natural number~$N$ as a product of natural numbers 
without regard to order corresponds to a thermally isolated non-interacting 
Bose gas in a one-dimensional potential with logarithmic energy eigenvalues.
This correspondence is used for characterising the probability distribution
which governs the number of factors in a randomly selected factorisation
of an asymptotically large~$N$. Asymptotic upper bounds on both the skewness 
and the excess of this distribution, and on the total number of factorisations,
are conjectured. The asymptotic formulas are checked against exact numerical 
data obtained with the help of recursion relations. It is also demonstrated 
that for large numbers which are the product of different primes the 
probability distribution approaches a Gaussian, while identical prime 
factors give rise to non-Gaussian statistics.
\end{abstract}




\maketitle

\section{Introduction}

Each integer number~$N$ can be written in a unique way as the product
of prime numbers~$p_i$ with integer multiplicities $n_i$,
\begin{equation}
   N = \prod_{i=1}^m \, p_i^{n_i}  \; .
\label{PFD}
\end{equation}
Every other possibility to decompose~$N$ into integer factors larger
than~$1$ is obtained by multiplying out some of its prime factors.
For $N = 30$, for instance, we have the factorisations
\begin{eqnarray}
   30 & = & 2 \cdot 3 \cdot 5   \nonumber \\
      & = & 5 \cdot 6           \nonumber \\
      & = & 3 \cdot 10          \nonumber \\
      & = & 2 \cdot 15          \nonumber \\
      & = & 30 \; .             
\end{eqnarray}
It is understood that the ordering of the factors does not matter here,
so that $5 \cdot 6$ is not distinguished from $6 \cdot 5$. Let $\Phi(N,k)$ 
denote the number of such factorisations of~$N$ which contain exactly 
$k$~factors, and let~$\Omega(N)$ be the total number of factorisations. 
Then, according to the above list, $\Phi(30,1) = 1$, $\Phi(30,2) = 3$, 
and $\Phi(30,3) = 1$, giving a total of~$\Omega(30) = 5$ factorisations. 

It is obvious that the number $\Omega(N)$ of possible factorisations
of~$N$ does not only depend on the total number of prime factors of~$N$, but 
also on the multiplicities with which the different prime factors occur: 
Taking $N = 12$, we have
\begin{eqnarray}
   12 & = & 2 \cdot 2 \cdot 3   \nonumber \\
      & = & 2 \cdot 6           \nonumber \\
      & = & 3 \cdot 4           \nonumber \\
      & = & 12 \; ,             
\end{eqnarray}
so that $\Phi(12,1) = 1$, $\Phi(12,2) = 2$, and $\Phi(12,3) = 1$, adding 
up to $\Omega(12) = 4$. Since two of the three prime factors of~$N = 12$
are equal, the number of different combinations of prime factors is less 
than for $N = 30$, which possesses three different prime factors. 

We therefore divide the natural numbers into equivalence classes: Two
numbers $N_1$ and $N_2$ are said to be equivalent if they give rise to
the same pattern of factorisations, meaning that $\Phi(N_1,k) = \Phi(N_2,k)$ 
for all~$k$. Particular equi\-valence classes consist of numbers which are 
some power of a single prime~$p$: If $N = p^m$, or 
\begin{equation}
   \ln N = m \, \ln p
\end{equation}
for some integer~$m$, the task of factorising~$N$ is equivalent to the 
task of partitioning the exponent~$m$ into integer summands, {\em i.e.\/},  
to the famous number-partitioning problem {\em partitio numerorum\/} 
introduced by Euler~\cite{Euler11}. This problem, which in itself plays a 
significant role in several areas of modern 
mathematics~\cite{HardyRamanujan18,Andrews98,AhlgrenOno01},
is connected to a number of topics occuring in statistical physics, ranging 
from lattice animals~\cite{WuEtAl96,BhatiaEtAl97} and combinatorial
optimisation~\cite{Mertens98} over Fermion-Boson transmutation~\cite{SM96} 
to quantum entropy and energy currents flowing in a one-dimensional
channel connecting thermal reservoirs~\cite{Blencowe01}.

In the general case, taking the logarithm of the prime factor 
decomposition~(\ref{PFD}) yields
\begin{equation}
   \ln N = \sum_{i = 1}^m n_i \, \ln p_i \; .
\label{LOG}
\end{equation}
Seen from the viewpoint of statistical physics, this latter equation allows 
for an interesting interpretation: If we consider an ideal Bose gas consisting 
of sufficiently many particles, confined such that the single-particle 
energies $\varepsilon_\nu$, when suitably made dimensionless, are  given 
by the logarithms of the primes, $\varepsilon_\nu = \ln p_\nu$ with
$p_\nu = 1,2,3,5,7,11,\ldots$ for $\nu = 0,1,2,3,4,5,\ldots\,$, the 
equation~(\ref{LOG}) indicates one particular microstate of the system, 
where the total energy $\ln N$ is distributed among the particles in such 
a way that $n_i$ is the occupation number of the energy level~$\varepsilon_i$. 
Additional particles which are not excited remain in the ground state, 
forming a Bose--Einstein condensate. Such surplus particles do not contribute
to the energy, since $\varepsilon_0 = \ln 1 = 0$. Ideal Bose gases with a 
single-particle spectrum corresponding to the sequence of the logarithms of 
the prime numbers have recently been studied by Tran and Bhaduri~\cite{Tran03},
with particular emphasis placed on differences between the fluctuation of the 
number of condensate particles in different statistical ensembles. 

A few remarks concerning low-dimensional, ideal Bose--Einstein condensates
might appear in order. The example of an ideal Bose gas of $M$ particles 
in a one-dimensional harmonic potential with oscillator frequency $\omega$
captures the essentials: If one considers the usual thermodynamic limit,
meaning $M \to \infty$ and $\omega \to 0$ such that the product $M\omega$
remains constant, the Bose--Einstein condensation temperature approaches 
zero~\cite{ZiffEtAl77}. However, this is not the relevant limit here;
we rather have to consider the limit of large particle numbers at 
{\em constant\/} oscillator frequency. In this case there still exists a 
nonzero characteristic temperature $\TC$ for the onset of a macroscopically 
large occupation of the ground state, roughly given by 
$\TC = (\hbar\omega/\kB) M/\ln(M)$, which approaches infinity for 
$M \to \infty$~\cite{KetterleVanDruten96,GH96}. Thus, there exists a condensate 
for any finite temperature, if the particle number is sufficiently large.  

Since we wish to treat {\em all\/} possible factorisations (not only those 
into primes) of a given number~$N$ without regard to the order of the factors, 
the Bose-gas analogy requires the single-particle spectrum
\begin{equation}
   \varepsilon_\nu = \ln(\nu + 1) \; , 
   \qquad \nu = 0,1,2,\ldots \; .
\end{equation} 
We point out that such a spectrum might actually be realisable: 
As shown in the appendix~\ref{ap:log}, within the quasi-classical 
approximation the eigenvalues of a particle in a one-dimensional
logarithmic potential $V(x) = V_0 \ln(|x|/L)$ are given by
$V_0\ln(2\nu + 1)$, up to a constant; the restriction to 
odd numbers $2\nu + 1$ is not essential.

The key point here is that the analogy between the factorisation problem
and an ideal Bose gas with logarithmic single-particle spectrum allows
us to invoke well-established methods from statistical physics for
obtaining information on number-theoretical properties of large composite
integers. In this paper, we focus on the probability distribution
\begin{equation}  
   P_N(k) = \frac{\Phi(N,k)}{\Omega(N)}
\label{DIS} 
\end{equation}
for given (asymptotically) large integers~$N$, {\em i.e.\/}, on the 
probability of finding~$k$ factors in a randomly selected factorisation
of a large~$N$. We will proceed as follows: In section~\ref{sec:rec}
we state recursion relations required for the numerical evaluation
of the exact quantities $\Phi(N,k)$, deferring the derivation to
appendix~\ref{ap:rec}. We then briefly explain in section~\ref{sec:asI}
the method used to obtain, by means of a detour from the microcanonical
to the canonical ensemble and back, asymptotic expressions for the cumulants 
of the distributions~(\ref{DIS}). In the following, we focus on the two 
extreme kinds of equivalence classes, namely those made up of powers of a 
single prime on the one hand, and those consisting of products of different 
primes on the other, and state the respective asymptotic formulas for the 
cumulants in sections~\ref{sec:asI} and~\ref{sec:asII}. In this way, an 
interesting feature will become apparent: While the presence of identical 
prime factors in the first case introduces Bose-like correlations which 
prevent the distributions~(\ref{DIS}) from becoming Gaussian even in the 
asymptotic limit, large products of different primes do indeed lead to 
almost Gaussian distributions. The paper closes with a brief summary in
section~\ref{sec:dis}.    

In passing, we point out that the problem considered in this paper should 
be clearly distinguished from a similar problem known as {\em factorisatio 
numerorum\/}, first investigated by Kalm\'ar~\cite{Kalmar30}. In this latter 
connection one counts all {\em ordered\/} sequences ($n_1, n_2, \ldots, 
n_k$) of integers $n_1, n_2, \ldots, n_k \ge 2$ which yield~$N$ when 
multiplied, $n_1 \cdot n_2 \cdot \ldots \cdot n_k = N$. Denoting the number 
of such ordered sequences by $a_N$ (with $a_1 = 1$), one deduces 
$ \sum_{N\ge1} a_N N^{-s} = [ 2-\zeta(s) ]^{-1} $, where $\zeta(s)$ is 
Riemann's zeta function; the task then is to find the asymptotic behaviour 
of the sum function
\begin{equation}
  A(x) \equiv \sum_{1\le N \le x} a_N \; .
\end{equation}
Recent results, and further information on this problem, have been collected 
in ref.~\cite{Hwang00}. In contrast, in the present paper we do {\em not\/} 
count different orderings of factors as different configurations, or 
microstates. It is precisely this identification of different orderings 
which leads to the Bose-gas analogy (with Bosons, the question ``which 
particle occupies which state'' is meaningless), and thus opens up the 
avenue followed here.

\section{\label{sec:rec}Recursion Relations}

In order to determine the exact distributions~(\ref{DIS}), we employ 
a recursion relation: Let $\Gamma_k({N})$ be the number of factorisations
of $N$ into~$k$ or less factors. Then, as shown in appendix~\ref{ap:rec},
one has  
\begin{eqnarray}
\label{eq:recpro}
   \Gamma_k({N}) & = & \frac{1}{k} \sum_{n=1}^k
   \sum_{\nu=1 \atop N \bmod {\nu}^n =0}^N
   {\Gamma}_{k-n}\left({N}/{\nu}^n\right) \; ,
\\ \nonumber
   \mbox{with} && \Gamma_k(1)=1\quad\mbox{and}\quad \Gamma_0({N}>1)=0 \; .
\end{eqnarray}
Contributions to the second sum in equation~(\ref{eq:recpro}) arise only 
when $\nu^n$ divides $N$. Starting from~$\Gamma_1(N) = 1$, the sequence 
$\Gamma_k(N)$ increases with increasing~$k$ (unless $N$ happens to be a prime) 
until it reaches its final value $\Omega(N)$, since~$N$ cannot be expressed 
as a product of more than~$\log_2(N)$ integer factors greater than~$1$:
\begin{equation}
   \Omega(N) = \Gamma_k(N) \; , \quad k \ge\log_2({N}) \; .
\end{equation}

\begin{figure}
\centerline{\includegraphics[width = 0.7\linewidth]{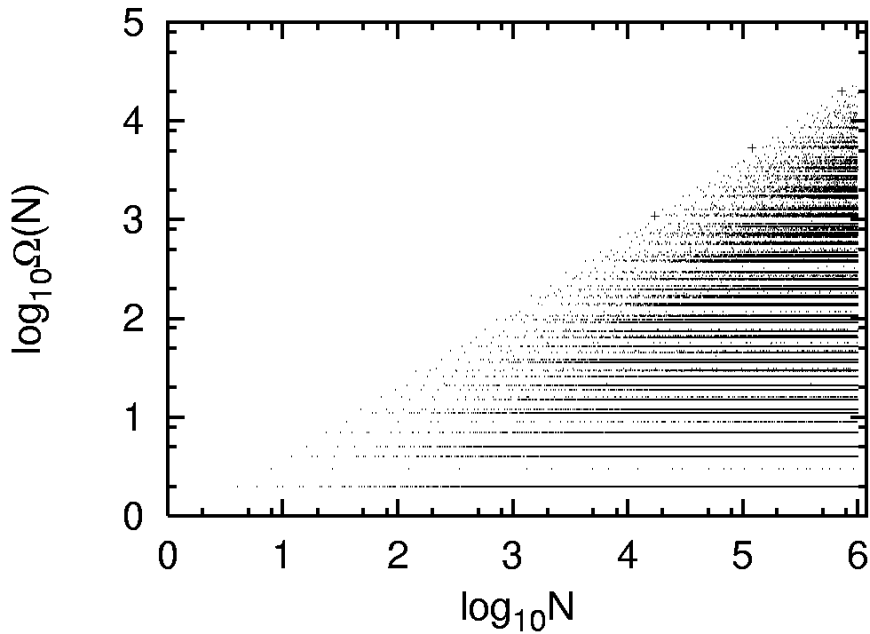}}


\caption[FIG.~1]{The number of possibilities $\Omega(N)$ to factorise an 
  integer~$N$ into products of natural numbers, for $ 1 \le N \le 10^6$.  
  For each $N$, the exact value $\Omega(N)$ has been computed with the
  help of the recursion relation~(\ref{eq:recpro}). While there are lots
  of numbers which are products of two primes, giving $\Omega({N})=2$, 
  only relatively few numbers can be written as the third power of a prime, 
  implying $\Omega({N}) = 3$. Composite numbers like 
  $17280  = 2^7 \cdot 3^3 \cdot 5 \,$, 
  $120960 = 2^7 \cdot 3^3 \cdot 5 \cdot7 \,$, or
  $725760 = 2^8 \cdot 3^4 \cdot 5 \cdot7 \,$ 
  yield rather high values of $\Omega({N})$, indicated by crosses.}  
\label{fig:omega}
\end{figure}

Figure~\ref{fig:omega} shows a logarithmic plot of $\Omega(N)$ for 
$1 \le N \le 10^6$. Each data point $[N, \Omega(N)]$ has been marked
by an individual dot; the equivalence classes clearly manifest themselves 
as horizontal lines. While these data might suggest an upper bound on 
$\Omega(N)$ on the order of $N^{0.77}$, the reader should be warned that 
the true asymptocis are not reached for $N$ as small as $10^6$; a correct 
upper bound will be stated later.

The number of possibilities to factorise~$N$ into exactly~$k$ factors 
is now given by
\begin{equation}
   \Phi(N,k) = \Gamma_k(N) - \Gamma_{k-1}(N) \; ,
\label{eq:dist}
\end{equation}
so that we have access to the probability distribution~(\ref{DIS}). As in 
previous investigations of ground-state fluctuations of non-interacting
and weakly interacting Bose gases~\cite{Scully00a,Scully00b,Weiss02}, 
it is useful to characterise such a distribution in terms of its 
cumulants~$\kappa^{(\ell)}$~\cite{Abramowitz72}, which directly quantify 
its deviation from a Gaussian: For a Gaussian distribution, 
$\kappa^{(\ell)} = 0$ for $\ell > 2$. The lowest four cumulants 
are related to the mean value $\langle k \rangle$ and the central moments 
$\mu^{(\ell)} = \langle(k - \langle k\rangle)^\ell\rangle$ through the 
relations~\cite{Abramowitz72}
\begin{eqnarray}
   \kappa^{(1)} & = & \langle k \rangle \; ,
\nonumber \\
   \kappa^{(2)} & = & \mu^{(2)} \; ,
\nonumber \\
   \kappa^{(3)} & = & \mu^{(3)} \; ,
\nonumber \\
   \kappa^{(4)} & = & \mu^{(4)} - 3 \left(\mu^{(2)}\right)^2 \; .
\end{eqnarray}
Normalising the third and the fourth cumulant we respect to the second,
one obtains the skewness
\begin{equation}
\label{eq:skewdef}
   \gamma_1 \equiv 
   \frac{\kappa^{(3)}}{\left(\kappa^{(2)}\right)^{3/2}} \; , 
\end{equation}   
which characterises the asymmetry of the underlying probability
distribution, and the excess
\begin{equation}   
\label{eq:exdef}   
   \gamma_2 \equiv 
   \frac{\kappa^{(4)}}{\left(\kappa^{(2)}\right)^{2}} \; ,
\end{equation} 
which characterises its flatness.

\begin{figure}
\centerline{\includegraphics[width = 0.7\linewidth]{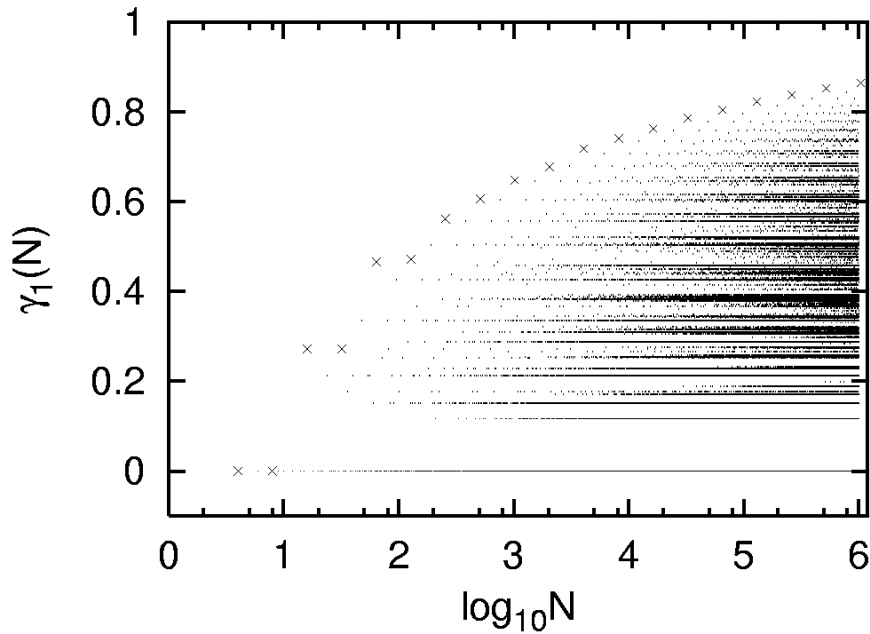}}


\caption[FIG.~2]{Exact skewness $\gamma_1(N)$, as defined in 
   equation~(\ref{eq:skewdef}), for the probability distributions~(\ref{DIS}),
   with $1 \le N \le 10^6$. There are some numbers~$N$ for which the 
   distribution is symmetric, so that $\gamma_1(N) = 0$. In contrast, 
   numbers which are integer powers of~$2$ give rise to a particularly 
   large skewness, as indicated by the crosses.} 
\label{fig:skew}
\end{figure}

In figure~\ref{fig:skew} we display exact data for the 
skewness~$\gamma_1({N})$ of the distributions~(\ref{DIS}), revealing that
their asymmetry becomes particularly large when $N = 2^m$. It is clear that
all numbers within the same equivalence class yield the same cumulants, and
thus the same values of $\gamma_1$ and $\gamma_2$.

Since the general algorithm~(\ref{eq:recpro}) requires rather a large amount 
of computer memory, it pays to focus on particular equivalence classes: 
For numbers~$N$ which are powers of a prime~$p$,
\begin{equation}
   N = p^m \; ,
\label{TY1}
\end{equation}
one has
\begin{equation}
  \Phi(p^m,k) = \sum_{\nu=1}^{\min\{m-k,k\}} \Phi(p^{m-k},\nu) \; ;
\label{eq:recp}
\end{equation} 
this recursion relation still has to be solved numerically.
If, on the other hand, $N$ is a product of distinct primes, 
\begin{equation}
   N = \prod_{i=1}^m \, p_i \; ,
   \qquad p_i \ne p_j \; \mbox{for} \; i \ne j \; , 
\label{DPF}
\end{equation}
the probability distribution is given by the relation
\begin{equation}
\label{eq:recdp}
  \Phi\!\left(\mbox{$\prod_{j=1}^{m}p_j$}\,,\;k\right)
   = \Phi\!\left(\mbox{$\prod_{j=1}^{m-1}p_j$}\,,\;k\!-\!1\right)
   +k\Phi\!\left(\mbox{$\prod_{j=1}^{m-1}p_j$}\,,\;k\right)
\end{equation}
 with
$\Phi(p_1,1) = 1 $
and
$ \Phi\!\left(\mbox{$\prod_{j=1}^{m-1}p_j$}\,,\;m\right) = 0 $.
As explained in appendix~\ref{ap:an}, this relation even admits a closed
solution:
\begin{equation}
   \Phi\!\left(\mbox{$\prod_{j=1}^{m}p_j$}\,,\;k\right)
   = \frac{(-1)^k}{k!}\sum_{\ell=1}^{k}{(-1)^{\ell}
   {k \choose \ell}{\ell}^m} \; .
\end{equation}

\section{\label{sec:asI} Asymptotic formulas for factorising powers
of a prime}

The problem of finding all factorisations of a given number $N$ 
corresponds, within the Bose-gas analogy, to a problem of microcanonical
statistics: Given the total energy $\ln N$, the task is to find all
possible microstates, {\em i.e.\/}, all possibilities for distributing
the energy over the available single-particle levels $\varepsilon_\nu$. 
Particles which are not excited and thus remain in the condensate 
contribute with $\ln \varepsilon_0 = 0$ to the energy, or with a factor 
of~$1$ to the product. Hence, the picture to have in mind is that of a 
partially condensed Bose gas, with the excited particles corresponding to 
factors larger than $1$, and a sufficiently large supply of condensed 
particles in the ground state. 

While the {\em microcanonical\/} requirement to keep the total energy constant
introduces severe technical difficulties, the statistics of partially
condensed ideal Bose gases become much simpler in the {\em canonical\/} 
ensemble, where the system is kept in contact with a thermal reservoir of 
temperature~$T$. It is precisely at this point that the Bose-gas picture
becomes essential: One can obtain information about the microcanonical number
factorisation problem by invoking the notion of temperature.    

We employ the first energy gap $\varepsilon_1 - \varepsilon_0$
to introduce the dimensionless inverse temperature
\begin{equation}
   b = (\varepsilon_1 - \varepsilon_0)\beta \; ,
\label{DEB}
\end{equation}  
where $\beta = 1/(\kB T)$, as usual. A decisive role for the canonical
statistics of a partially condensed, ideal Bose gas is played by the grand 
canonical partition function $\Xi_{\rm ex}(b,z)$ of an auxiliary system from 
which the single-particle ground state $\varepsilon_0$ has been removed, 
while all other levels $\varepsilon_\nu$ remain unaltered. If there is an
infinite supply of condensed particles, in the spirit of the so-called
Maxwell's demon ensemble~\cite{Navez97}, this function has the exact 
integral representation~\cite{HolthausEtAl98,HolthausEtAl01}   
\begin{equation}
\label{LNX}
   \ln \Xi_{\rm ex}(b,z) =
   \frac{1}{2\pi i} \int_{\tau-i\infty}^{\tau+i\infty} \! \rd t \;
   b^{-t} \, \Gamma(t) \, \eta(t) \, g_{t+1}(z) \; , 
\end{equation}   
where
\begin{equation}
   g_\alpha(z) = \sum_{n=1}^\infty \frac{z^n}{n^\alpha}
\end{equation}   
denotes the familiar Bose functions~\cite{Pathria96}, and 
\begin{equation}
\label{eq:etadef}
   \eta(t) \equiv \sum_{\nu=1}^\infty 
   \left(\frac{\varepsilon_1-\varepsilon_0}
              {\varepsilon_\nu-\varepsilon_0}\right)^t
\end{equation}
is a series determined by the single-particle spectrum. The real 
number $\tau > 0$ in equation~(\ref{LNX}) is chosen such that all poles of 
the integrand lie on the left hand side of the path of integration. The
value of this representation~(\ref{LNX}), and of the subsequent 
representation~(\ref{eq:cancum}), lies in the fact that it lends itself
to the derivation of an asymptotic series which captures the 
low-$b$-behaviour, if one collects, from right to left on the real axis,
the residues of the respective integrands. With respect to an actual Bose gas, 
the underlying assumption of an infinite number of ground-state particles 
restricts the validity of equation~(\ref{LNX}) to temperatures low enough to 
guarantee the existence of a condensate. In the case of the number 
factorisation problem there is no similar restriction, owing to the fact 
that multiplying any product by an arbitrary amount of factors of unity 
does not change its value: In the number-theoretic context the 
representation~(\ref{LNX}) is {\em exact\/}. 
  
The $\ell$-th cumulants $\kappa_{\rm cn}^{(\ell)}(b)$ of the canonical 
distribution governing the number of excited particles in the gas are then 
immediately obtained by differentiation, 
\begin{equation}  
   \kappa_{\rm cn}^{(\ell)}(b) = \left. 
   \left(z \frac{\partial}{\partial z}\right)^\ell
   \ln \Xi_{\rm ex}(b,z) \right|_{z=1} \; ,
\label{KAP}
\end{equation}
giving~\cite{HolthausEtAl98,HolthausEtAl01}
\begin{equation}
   \kappa_{\rm cn}^{(\ell)}(b) =
   \frac{1}{2\pi i} \int_{\tau-i\infty}^{\tau+i\infty} \! \rd t \;
   b^{-t} \, \Gamma(t) \, \eta(t) \, \zeta(t+1-\ell) \; ,
\label{eq:cancum}
\end{equation} 
where~$\zeta(t)=\sum_{n=1}^{\infty}n^{-t}$ is the Riemann zeta function.
We will only consider non-interacting Bosons in trapping potentials with 
single particle energy levels~$\varepsilon_{\nu}$ ($\nu=0,1,2\ldots$) such 
that the series~(\ref{eq:etadef}) converges for $t>t_0$, with some real 
$t_0 > 0$.

The strategy of employing the canonical ensemble for solving a microcanonical 
problem hinges on the possibility to get rid of the temperature in a second 
step, and to find an expression for the microcanonical cumulants. This is 
done with the help of the saddle-point method~\cite{Navez97}: The 
``energy-temperature'' relation reads
\begin{equation}
\label{eq:en_temp}
   n + 1 = -\left.\frac{\partial}{\partial b}
   \ln\Xi_{\rm ex}(b,z)\right|_{b=b_0(z)} \; ,
\end{equation}
where~$n$ denotes the energy in units of $\varepsilon_1-\varepsilon_0$, and 
$b_0(z)$ is the saddle point. The generating function of the microcanonical 
cumulants then takes the form~\cite{Weiss03} 
\begin{eqnarray}
  \ln Y(n,z) & = & \ln \Xi_{\rm ex}(b_0(z),z) + n b_0(z) 
  - \frac{1}{2}\ln 2\pi
\nonumber \\ & & 
  - \frac{1}{2}\ln \! \left.\left(\frac{\partial^2}{\partial b^2}
  \ln \Xi_{\rm ex}(b,z) \right)\right|_{b=b_0(z)} \; , 
\label{eq:micgen}   
\end{eqnarray}
and the desired microcanonical cumulants are finally calculated by  
taking derivatives,
\begin{equation}
  \kappa_{\rm mc}^{(\ell)}(n) = \left. \left( z\frac{\rd}{\rd z}\right)^\ell 
  \ln Y(n,z) \right|_{z=1} \; .     
\label{MCC}
\end{equation}

The method sketched here requires knowledge about the analytical properties
of the function~$\eta(t)$ introduced in equation~(\ref{eq:etadef}). For 
applications to the number factorisation problem, the single-particle
energies are determined by the equivalence class of the number~$N$: 
In the simplest case~(\ref{TY1}), the possible single-particle energies are 
integer multiples of $\ln p$, so that $\eta(t)$ coincides with the Riemann 
zeta function $\zeta(t)$, and the factorisation of~$N$ is equivalent to the 
partition of $m = \log_p N$ into summands. The evaluation of the above 
formalism for the partition problem has already been discussed in 
detail~\cite{Weiss03}, so we merely quote the results: For numbers of the 
type~(\ref{TY1}), the asymptotic formula for the skewness~(\ref{eq:skewdef}) 
reads
\begin{eqnarray}
   \gamma_1 & = & \frac{12\sqrt{6}\,\zeta(3)}{\pi^3}
\nonumber \\
   & + &  \frac{1}{\sqrt{\log_p(N)}}
          \left[ 0.10128 \left[\ln(\log_p(N))\right]^2 
        - 0.37376 \, \ln(\log_p(N)) - 1.7078 \right]
\nonumber \\
   & + & \frac{1}{\log_p(N)}\left[ 0.0075008 \left[\ln(\log_p(N))\right]^4 
        + 0.025681 \left[\ln(\log_p(N))\right]^3 \right.
\nonumber \\
   & + & \left. 0.020024 
  \left[\ln(\log_p(N))\right]^2 - 0.23028 \, \ln(\log_p(N)) - 0.56984 \right] 
\nonumber \\    
   & + & {\mathcal O}(\log_p(N)^{-3/2}) \; , 
\label{eq:g1as}                  
\end{eqnarray}
while the excess~(\ref{eq:exdef}) takes the form
\begin{eqnarray}
   \gamma_2 & = & 2.4
\nonumber \\ 
   & + & \frac{1}{\sqrt{\log_p(N)}}
         \left[ 0.28440 \left[\ln(\log_p(N))\right]^2 
       - 0.56714 \, \ln(\log_p(N)) - 10.064 \right]
\nonumber \\
   & + & \frac{1}{\log_p(N)}\left[ 0.025276 \left[\ln(\log_p(N))\right]^4 
        + 0.022329 \left[\ln(\log_p(N))\right]^3 \right.
\nonumber \\ & - &  \left.
         0.33809 \left[\ln(\log_p(N))\right]^2 + 0.73538 \, \ln(\log_p(N)) 
         + 3.7863 \right] 
\nonumber \\    
   & + & {\mathcal O}(\log_p(N)^{-3/2}) \; . 
\label{eq:g2as}   
\end{eqnarray}

\begin{figure}
\centerline{\includegraphics[width = 0.6\linewidth]{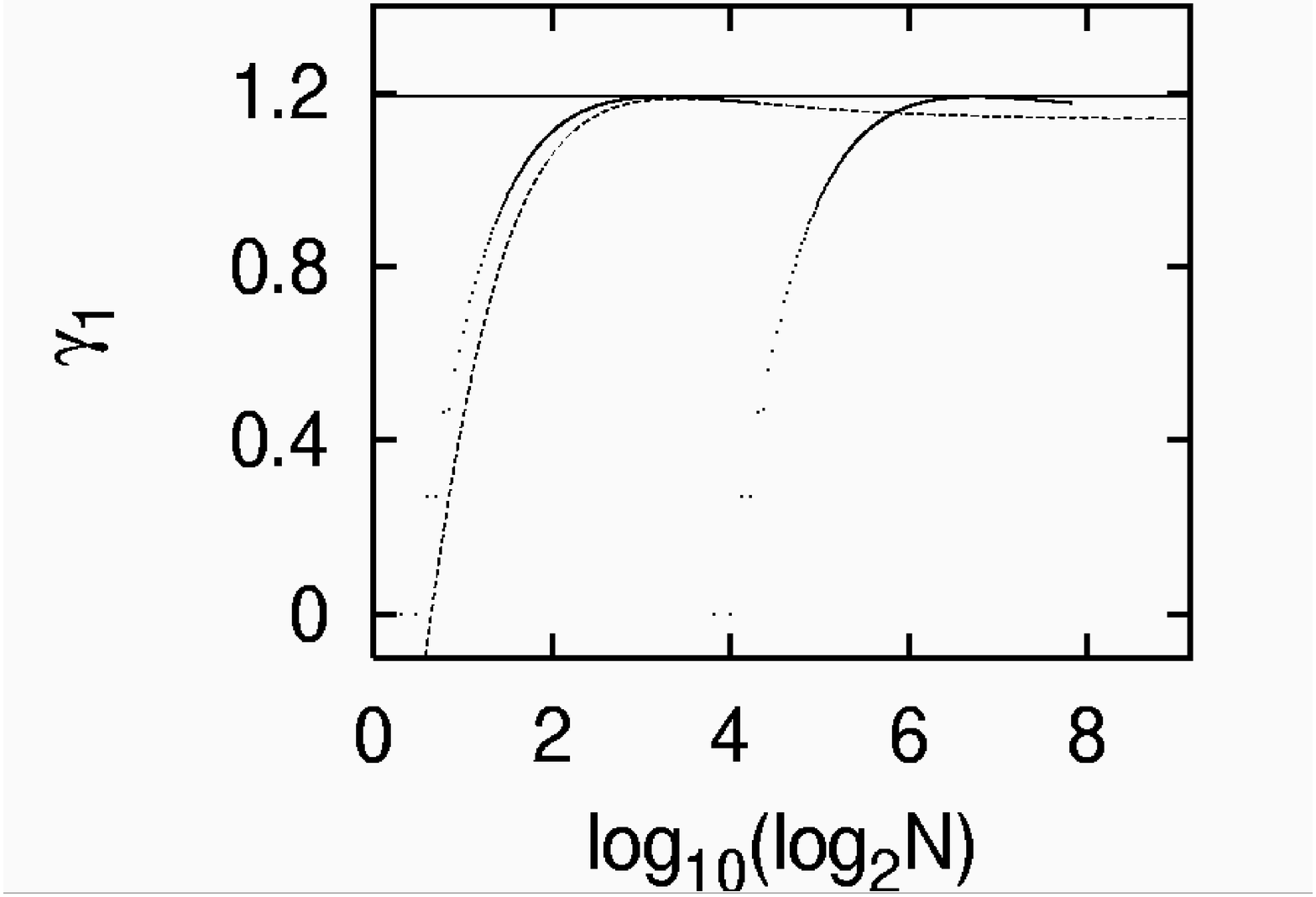}}


\caption[FIG.~3]{Skewness~(\ref{eq:skewdef}) for large numbers $N = 2^m$ 
   and $N = (10^{1000} + 453)^m$, as obtained from the recursion 
   relation~(\ref{eq:recp}) (dots). The dashes correspond to the evaluation 
   of the asymptotic formula~(\ref{eq:g1as}) for $N = 2^m$; the horizontal 
   line indicates the maximum value $\gamma_{1,\;\rm max} \simeq 1.1907$.}
\label{fig:skewlarge}
\end{figure}

\begin{figure}
\centerline{\includegraphics[width = 0.6\linewidth]{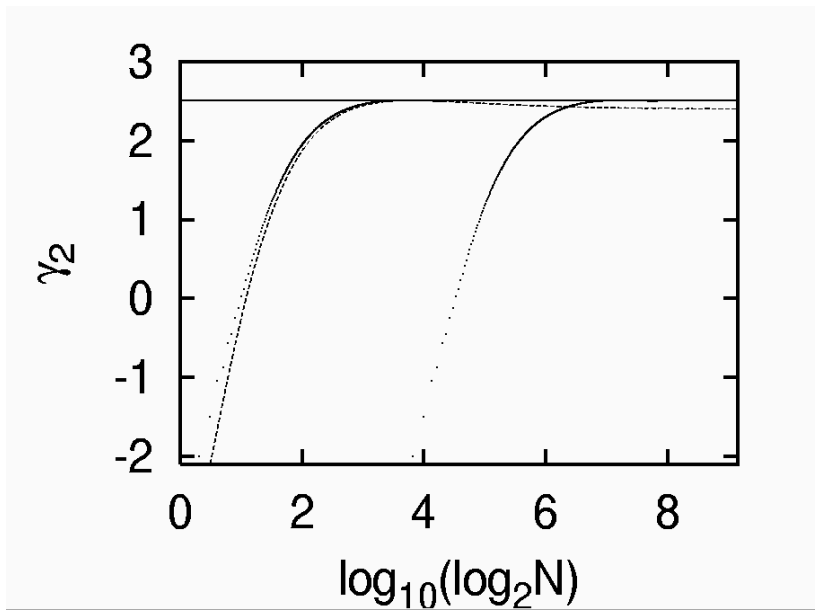}}


\caption[FIG.~4]{Excess~(\ref{eq:exdef}) for large numbers $N = 2^m$ 
   and $N = (10^{1000} + 453)^m$, as obtained from the recursion 
   relation~(\ref{eq:recp}) (dots). The dashes correspond to the evaluation 
   of the asymptotic formula~(\ref{eq:g2as}) for $N = 2^m$; the horizontal 
   line indicates the maximum value $\gamma_{2,\;\rm max} \simeq 2.5120$.}
\label{fig:exlarge}
\end{figure}

In order to demonstrate the accuracy of these formulas, we compare in
figures~\ref{fig:skewlarge} and \ref{fig:exlarge} exact values for the 
skewness and the excess, corresponding to large numbers $N = p^m$, 
with the predictions of the equations~(\ref{eq:g1as}) and (\ref{eq:g2as}). 
The agreement is excellent. It is also visible that both $\gamma_1(N)$ 
and $\gamma_2(N)$ approach constant values only for numbers~$N$ of the 
order~$p^{\left(10^{10}\right)}$. The finding that the limiting values of 
skewness and excess are nonzero expresses the fact that for numbers of the
form~(\ref{TY1}) the distribution~(\ref{DIS}) remains non-Gaussian even 
for asymptotically large~$N$. 

The exact numerical data further reveal that both skewness and excess reach 
their limiting values from above; the skewness adopts its maximum value
\begin{equation}
\label{eq:g1max}
   \gamma_{1,\;\rm max} \simeq
   1.1907
\end{equation}
for $N = p^{1730}$. The maximum of the excess,  
\begin{equation}
\label{eq:g2max}
   \gamma_{2,\;\rm max} \simeq
   2.5120
\end{equation}
lies at $N = p^{5507}$.

\section{\label{sec:asII} Asymptotic formulas for factorising products
of different primes}

We now turn to the equivalence classes consisting of products of 
different primes, and restrict ourselves to numbers~$N$ which are the 
product of the first $m$ primes,
\begin{equation}
   N = \prod_{i=1}^m \, p_i \; ,
\label{PRO}
\end{equation}
where $m$ is large. In this case, the single-particle energies 
accessible to the Bose gas are given by
\begin{equation}
   \varepsilon_{\{\nu_i\}_{i=1\ldots m}} =
   \sum_{i=1}^{m} \nu_i\ln(p_i) \; , \hspace*{1cm} \nu_i\in \{0,1\} \; .
\label{ORE}
\end{equation}
The restriction $\nu_i\in \{0,1\}$ stems from the fact that each prime
$p_i$ does either contribute to a given factor of $N$, or it does not.
The task then is to characterise the analytical properties of the
series $\eta(t)$ corresponding to this spectrum. We circumvent this 
severe difficulty by considering a simpler, but combinatorically 
equivalent problem: We replace the actual energy levels~(\ref{ORE}) by  
\begin{equation}
   \varepsilon_{\{\nu_i\}_{i=1\ldots m}} =
   \sum_{i=1}^{m} \nu_i \, a^{i-1} \; , \hspace*{1cm} \nu_i\in \{0,1\} \; ,
\label{eq:ena}
\end{equation}
where $a>1$ is a transcendental number which has to assure the ``prime''
character of $\exp(a^m)$ in the sense that it be not possible to express~$a^m$ 
as a sum of the form~$\sum_{i=1}^{m-1} n_i a^i$, for arbitrary integers
$n_i$, since a prime cannot be represented as a product of arbitrary
powers of lower primes, and any $m$, since the function $\eta(t)$ provided 
by the substitute~(\ref{eq:ena}) acquires a contribution for each~$m$. This 
requirement forces us to take the limit $a \to 1$ in the end: Otherwise, there 
would be an infinite sequence of integers $m$ for which the forbidden equality 
$a^m = \sum_{i=1}^{m-1} n_i a^i$ could be satisfied with arbitrary 
accuracy. It is also clear that the total energy~$\sum_{i=1}^{m}\ln(p_i)$ 
has to be replaced by
\begin{equation}
   n \equiv \sum_{i=1}^{m}a^{i-1} \; .
\label{eq:n}
\end{equation}
With this surrogate, we now have
\begin{equation}
   \eta(t) = \sum_{m=0}^{\infty}
   \sum_{ \left\{ \{\nu_i\}_{i=1\ldots m+1} \atop \nu_{m+1}=1 \right\} }
   \frac{1}{\left(\varepsilon_{\{\nu_i\}}\right)^t} \; ,
\end{equation}
and easily obtain upper and lower bounds on the second sum, assuming~$t>0$:
\begin{equation}
   \frac{2^{m}}{\left(a^m\right)^t}
   \ge
   \sum_{ \left\{ \{\nu_i\}_{i=1\ldots m+1} \atop \nu_{m+1}=1 \right\} }
   \frac1{\left(\varepsilon_{\{\nu_i\}}\right)^t} 
   \ge
   \frac{2^{m}}{\left(\frac{a^{m+1}-1}{a-1}\right)^t}
   >
   \frac{2^{m}}{\left(\frac{a^{m+1}}{a-1}\right)^t} \; .
\end{equation}
Thus, for $t>\frac{\ln(2)}{\ln(a)}$ we can perform the sum over~$m$ and
obtain 
\begin{equation}
   \frac1{1-\frac2{a^t}}
   > \eta(t) >
   \left(\frac{a-1}a\right)^t\frac1{1-\frac2{a^t}} \; ,
\end{equation}
implying that $\eta(t)$ has a simple pole at
\begin{equation}
\label{DEC}
   c \equiv \frac{\ln(2)}{\ln(a)} \; , 
\end{equation}
with a residuum~$r$ which can at least be estimated, 
\begin{equation}
   \label{eq:rboarders}
   \frac1{\ln{a}}
   > r >
   \frac{\left(\frac{a-1}{a}\right)^{\frac{\ln(2)}{\ln(a)}}}{\ln{a}}\;.
\end{equation}
These informations suffice for an approximate evaluation of the integral 
representation~(\ref{eq:cancum}) of the canonical cumulants. We are 
interested in their asymptotic behaviour for small~$b$, as defined 
in equation~(\ref{DEB}): While the temperature has to be low enough to 
guarantee the existence of a Bose condensate, $\kB T = 1/\beta$ 
has to remain large in comparison with $\varepsilon_1 - \varepsilon_0$, so 
that sufficiently many states above the ground state remain populated. Now 
the small-$b$-asymptotics of the cumulants~(\ref{eq:cancum}) are determined 
by the rightmost pole of the respective integrand. Since $\Gamma(t)$ has 
simple poles for $t = 0,-1,-2,\ldots\,$, and $\zeta(t+1-\ell)$ has a simple 
pole at $t = \ell$, the dominant pole is provided by $\eta(t)$ at $t = c\,$: 
Taking $a$ close to unity, as required, results in a value of~$c$ larger 
than any fixed, finite number. 

Restricting ourselves to this dominant pole, the residue theorem 
then yields
\begin{equation}
   \kappa_{\rm cn}^{(\ell)}(b) \sim 
   \frac{\Gamma(c) \, \zeta(c+1-\ell) \, r}{b^c} \; , 
   \qquad \ell = 0,1,2,\ldots \; ,
\end{equation}
with the $\sim$-sign indicating asymptotic equality. It follows 
immediately that
\begin{equation}
   \lim_{b\to 0} \frac{\kappa_{\rm cn}^{(\ell)}(b)}
   {\left(\kappa_{\rm cn}^{(2)}(b)\right)^{\frac{\ell}{2}}} = 0 \; ,
   \qquad \ell > 2 \; ,
\end{equation}
which means that the canonical distribution becomes Gaussian-like
for small temperatures. 

Next, one needs the energy-temperature relation~(\ref{eq:en_temp}) in 
order to return to the microcanonical ensemble. Since, according to 
equation~(\ref{KAP}), $\ln \Xi_{\rm ex}(b,1)$ coincides with 
$\kappa_{\rm cn}^{(0)}(b)$, to leading order this relation takes the form
\begin{equation}
  b(n) \sim \left(\frac{c \, \Gamma(c) \, \zeta (c+1) \, r}
                       {n+1}\right)^{\frac1{c+1}} \; ,
\end{equation}
and the leading-order term of the microcanonical cumulants~(\ref{MCC}) is 
given by
\begin{equation}
\label{eq:mclead}
   \kappa_{\rm mc}^{(\ell)}(n) \sim
   {\frac{\Gamma(c) \, \zeta(c+1-\ell) \, r}{\left[ b(n) \right]^c}} \;.
\end{equation}
Utilising
\begin{equation}
   \frac{\Gamma(c+1)^{1/(c+1)}}{c} \; \longrightarrow \; \exp(-1)
\end{equation}
for $c \to \infty$, we then find 
\begin{equation}
\label{eq:cummc}
   \kappa_{\rm mc}^{(\ell)}(n) \sim \exp(-1)(n+1) \; ,
   \qquad \ell \ge 1 \; .
\end{equation}
As an immediate consequence, we have
\begin{equation}
   \lim_{n\to\infty} \frac{\kappa_{\rm mc}^{(\ell)}(n)}
   {\left(\kappa_{\rm mc}^{(2)}(n)\right)^{\frac{\ell}{2}}} = 0 \; ,
   \qquad \ell > 2 \; ,
\end{equation}
so that the approach to a Gaussian is also met in the microcanonical
ensemble. 

We now have to get rid of the auxiliary energy-like quantity~$n$, 
and to re-introduce the number to be factorised,~$N$. By virtue of the
definition~(\ref{eq:n}) one has $n \sim m$ for $a \to 1$, meaning
that the ``energy'' approaches the number of prime factors of $N$
when~$N$ becomes large. Hence, we find    
\begin{equation}
  \gamma_1 \sim \frac1{\sqrt{\exp(-1)\, (m+1)}}
\end{equation}
and
\begin{equation}
  \gamma_2 \sim \frac1{{\exp(-1)\,(m+1)}} \; , 
\end{equation}
so that skewness and excess have been expressed in terms of the number~$m$ 
of primes contained in the product~(\ref{PRO}), provided $m$ is sufficiently 
large. The final links in our chain of arguments are then provided by two 
results from analytic number theory: Firstly, if $N$ equals the product of 
the first~$m$ primes from 2 to $p_m$, as in our case~(\ref{PRO}), 
one has~\cite{Rademacher73} 
\begin{equation}
   \ln N \sim p_m \; .
\end{equation}
Secondly, the number $\pi(p_m)$ of primes less than $p_m$, which is $m$, 
is asymptotically given by~\cite{Rademacher73}
\begin{equation}
   \pi(p_m) \sim \frac{p_m}{\ln\left(p_m \right)} \; .
\end{equation}
These estimates combine to yield
\begin{equation}
   m \sim \frac{\ln(N)}{\ln(\ln(N))} \; .
\label{MLN}
\end{equation}
Thus, the final asymptotic expressions for the skewness and the excess
of the distributions~(\ref{DIS}) pertaining to large numbers~$N$ 
expressable as a product of all primes up to some $p_m$ become   
\begin{equation}
\label{eq:g1asn}
   \gamma_1 \sim \frac{1}
   {\sqrt{\exp(-1)\left(\frac{\ln(N)}{\ln(\ln(N))}+1\right)}} 
\end{equation}
and 
\begin{equation}
\label{eq:g2asn}
  \gamma_2\sim \frac{1}
  {\exp(-1)\left(\frac{\ln(N)}{\ln(\ln(N))}+1\right)} \;.
\end{equation}

\begin{figure}
\centerline{\includegraphics[width = 0.6\linewidth,angle=-90]{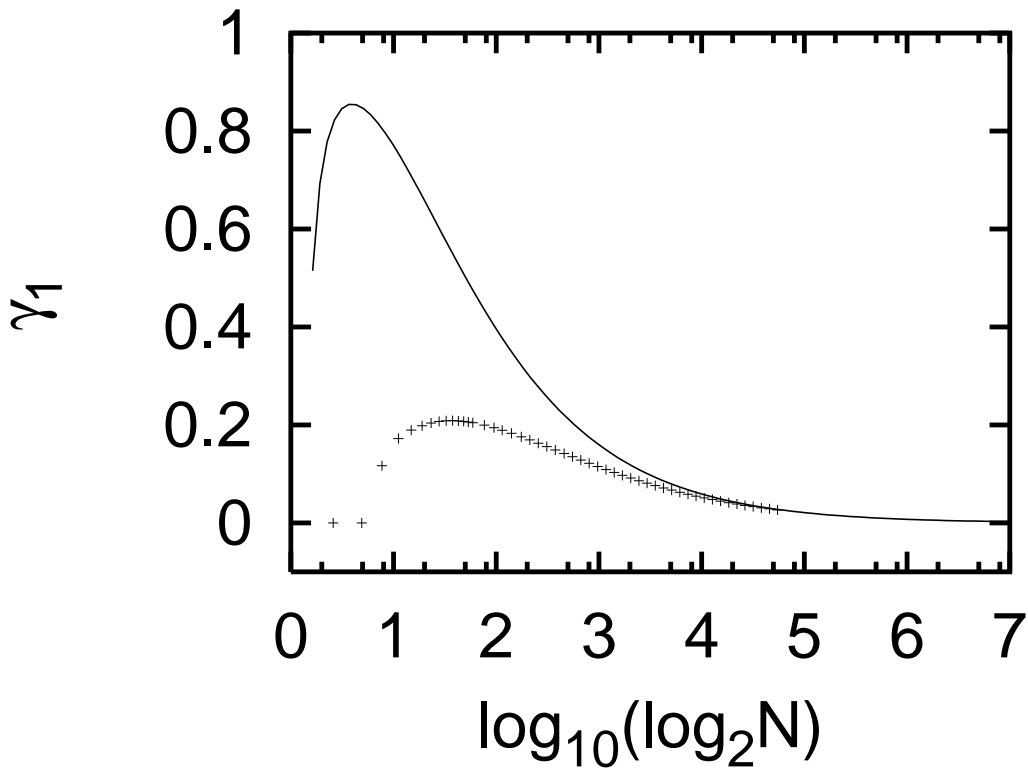}}


\caption[FIG.~5]{Skewness~(\ref{eq:skewdef}) for large numbers $N$ which
   are the product of the first $m$ primes. Crosses indicate exact
   numerical data, obtained with the recursion relation~(\ref{eq:recdp});
   the full line corresponds to the asymptotic expression~(\ref{eq:g1asn}).} 
\label{fig:skewprod}
\end{figure}

\begin{figure}
\centerline{\includegraphics[width = 0.6\linewidth,angle=-90]{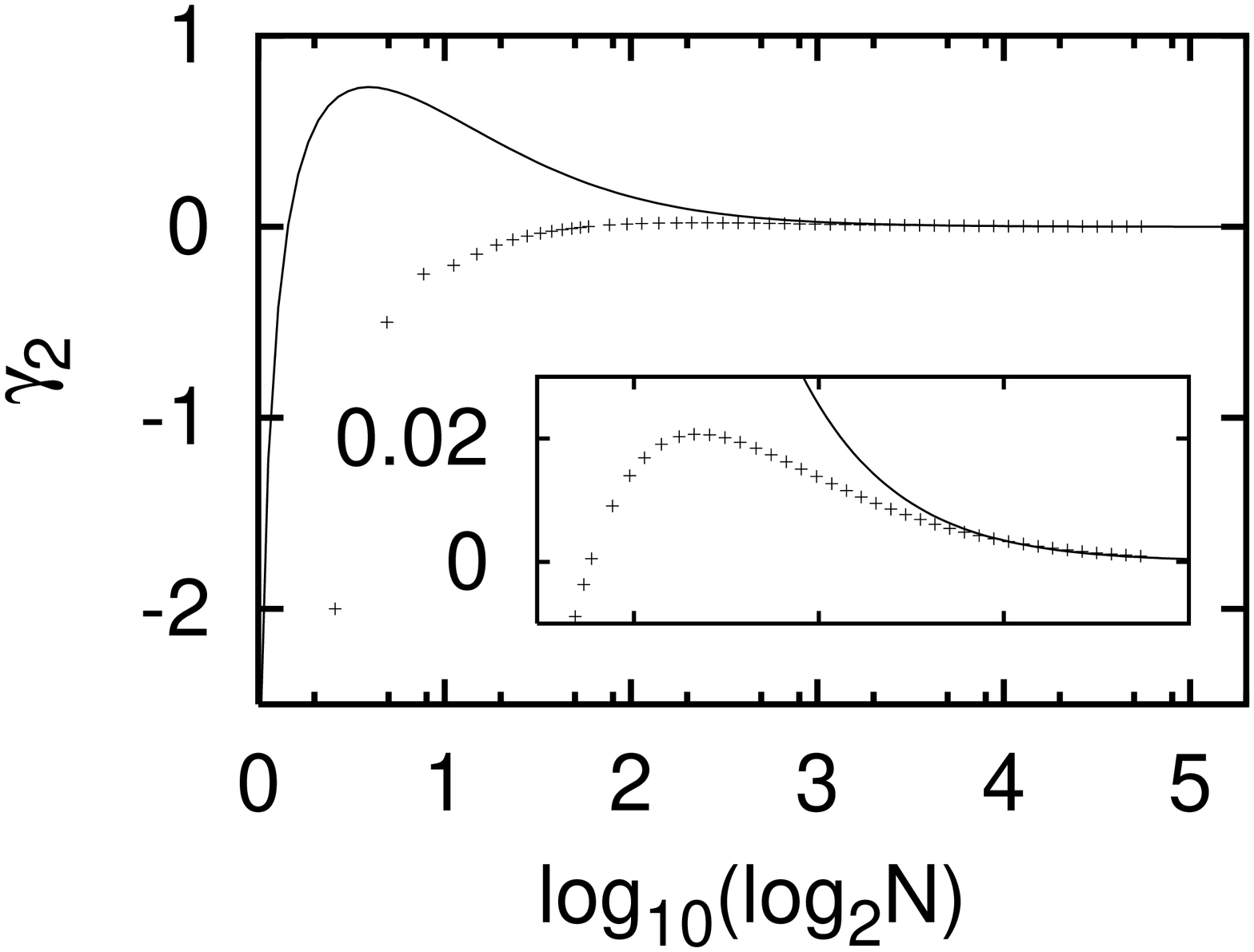}}


\caption[FIG.~6]{Excess~(\ref{eq:exdef}) for large numbers $N$ which 
   are the product of the first $m$ primes. Crosses indicate exact
   numerical data, obtained with the recursion relation~(\ref{eq:recdp});
   the full line corresponds to the asymptotic expression~(\ref{eq:g2asn}).
   The inset demonstrates that this expression actually describes the
   asymptotics correctly.} 
\label{fig:exprod}
\end{figure}

Again, we check these results against exact numerical calculations:
The figures~\ref{fig:skewprod} and~\ref{fig:exprod} depict skewness and
excess of the probability distributions~(\ref{DIS}) corresponding to
large numbers~$N$ of the type~(\ref{PRO}), again contrasting exact data 
points with the predictions of the asymptotic formulas. For smaller~$N$, 
the agreement is not quite as good as in the previous 
figures~\ref{fig:skewlarge} and \ref{fig:exlarge}. This can be attributed 
to the fact that equations~(\ref{eq:g1asn}) and~(\ref{eq:g2asn}) merely 
stem from a leading-order analysis, as necessitated by the rather 
complicated function~$\eta(t)$ underlying the~$N$ studied here. Nonetheless,
the approach to the Gaussian limits $\gamma_1 = 0$ and $\gamma_2 = 0$ is 
captured correctly.

\section{\label{sec:dis}Discussion}

We have used the correspondence between an ideal Bose gas with
logarithmic single-particle levels and the number factorisation problem
to characterise the ``number-of-factors''-distribution~(\ref{DIS}) for 
large integers~$N$. The properties of this distribution depend on the 
equivalence class of~$N$, that is, on the multiplicity of its prime
factors. The case~(\ref{TY1}), with all the prime factors being equal, 
constitutes one extreme; we have shown that for numbers of this type
the skewness~(\ref{eq:skewdef}) and the excess~(\ref{eq:exdef})
asymptotically approach the limiting values $\gamma_{1,\infty} =
12\sqrt{6}\,\zeta(3)/\pi^3 \simeq 1.1395$ and $\gamma_{2,\infty} = 2.4$.  
We conjecture that for {\em any\/} number $N$, the skewness $\gamma_1(N)$
and the excess $\gamma_2(N)$ do not exceed the maximum values 
(\ref{eq:g1max}) and (\ref{eq:g2max}) adopted by numbers of this particular
type:

\newtheorem{guess}{Conjecture}
\begin{guess}
   For every integer $N$, the skewness~(\ref{eq:skewdef}) of the probability 
   distribution~(\ref{DIS}) governing the number of factors in a randomly 
   selected product decomposition is bounded from above by
   $\gamma_{1,\;\rm max} \simeq 1.1906570491$,    
   while its excess~(\ref{eq:exdef}) is bounded by
   $\gamma_{2,\;\rm max} \simeq 2.5119565935$.
\end{guess}   
   
Moreover, for a given total number of prime factors (which equals 
$\sum_{i=1}^m n_i$ for $N=\prod_{i=1}^m p_i^{n_i}$), the total number of 
factorisations $\Omega(N)$ becomes largest for integers~$N$ of the 
type~(\ref{DPF}), for which all prime factors differ from each other. Upper 
and lower bounds on the number of factorisations for integers of this 
particular type have been derived in appendix~\ref{ap:an}. While the lower 
bound clearly pertains only to the equivalence classes considered there, 
we conjecture that the upper bound inferred from the inequality~(\ref{EST}) 
holds for {\em all\/} large integers. Again utilising the asymptotic 
relation~(\ref{MLN}), we thus formulate   

\begin{guess}
\label{CO2}
   For large integers~$N$, an upper bound on the number~$\Omega(N)$ of
   possible factorisations is provided by 
\begin{eqnarray}
   \ln\left[\Omega(N)\right]
    & < & \frac{\ln(N)}{\ln (\ln(N))}
      \ln\left[\frac{\ln(N)}{\ln(\ln(N))}\right]
\nonumber \\ & \sim & 
   \ln(N) \, .
\label{eq:omegabound}
\end{eqnarray}
\end{guess}

\begin{figure}

\centerline{\includegraphics[width = 0.6\linewidth,angle=-90]{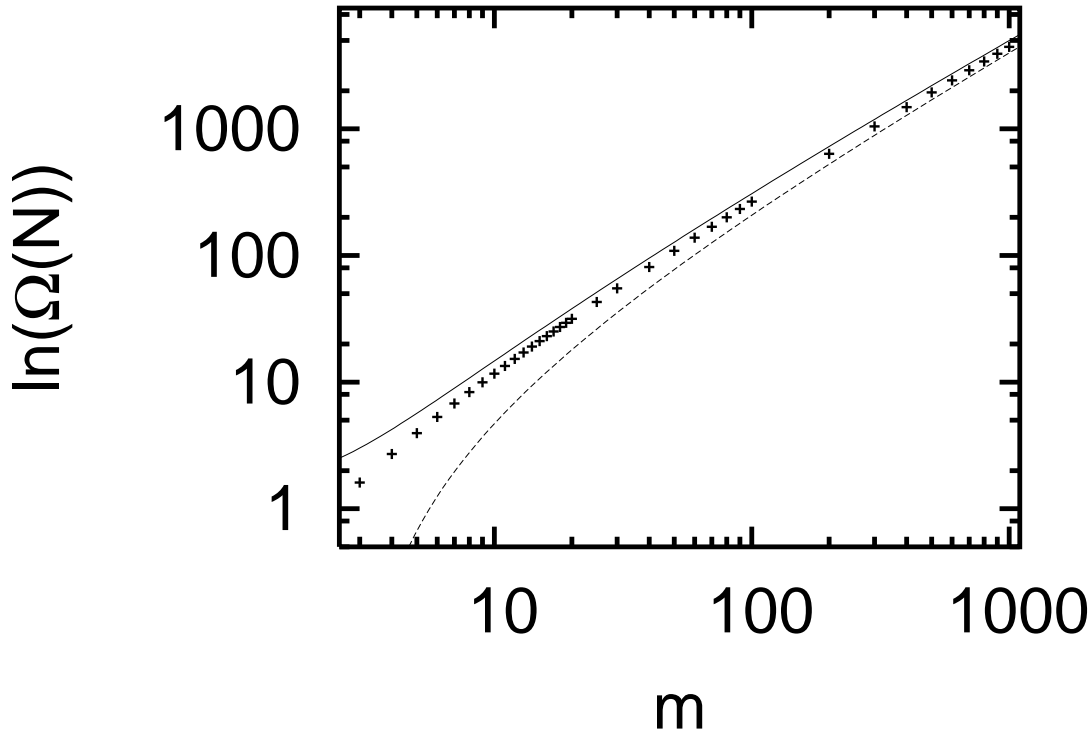}}


\caption[FIG.~7]{Upper and lower bound~(\ref{EST}) for the logarithm of
   the number of factorisations of large integers~$N$ which are the
   products of the first~$m$ primes, in comparison with exact numerical
   data (crosses). It follows that an asymptotic upper bound on
   $\ln \Omega(N)$ behaves at least as $\ln N$.}
\label{fig:bounds}
\end{figure}

In figure~\ref{fig:bounds} we display the bounds~(\ref{EST}) for products 
of different primes, together with exact numerical data. Despite the somewhat 
crude approximations, these bounds describe the data quite well. Hence, when 
$\ln(\ln(N)) \gg 1$, the number $\Omega(N)$ grows as~$N$ at least for some 
integers~$N$, which was not obvious at all from the limited data collected 
in figure~\ref{fig:omega}.

The actual content of conjecture~\ref{CO2} lies in the circumstance that 
it might still be possible to trade equality of prime factors for  
smallness of~$N$: Constructing some composite integer by choosing more 
than one prime factor equal to $p_1 = 2$, say, certainly reduces the value 
of $\Omega$ below the one that is attained when all prime factors are
different, but also reduces the value of $N$ itself. In this way, one
might try and maintain a relatively high value of $\Omega(N)$, while
minimising~$N$. Indeed, the three examples singled out by the crosses in 
figure~\ref{fig:omega} indicate that in some cases such a procedure might 
lead to data points which fall at least close to the envelope of all pairs 
$[N, \Omega(N)]$. However, we conjecture that the bound~(\ref{eq:omegabound}) 
holds nonetheless.  

For powers of a single prime~$p$, the factorisation problem is 
equivalent to Euler's number partitioning problem~\cite{Euler11}, 
so that the asymptotics can be deduced from the Hardy-Ramanujan 
formula~\cite{HardyRamanujan18} for the number of partitions: 
\begin{equation}
   \ln[\Omega(N)] \sim \pi \sqrt{\frac{2}{3} \log_{p} N}
   \qquad \mbox{for} \quad N=p^m \;.
\end{equation}
As expected, in this case $\ln[\Omega(N)]$ lies well below the conjectured 
upper bound~(\ref{eq:omegabound}). On the other hand, a strict upper bound 
can be established as follows: $N$ cannot be the product of more than 
$\log_2(N)$ primes. Assuming that these are all different (which can only 
overestimate the number of factors), and again using the result~(\ref{EST}), 
one finds
\begin{equation}
   \ln[\Omega(N)] < \log_2(N) \, \ln[\log_2(N)]
   \qquad \mbox{for all~$N$} \; .
\end{equation}
Our conjectured bound~(\ref{eq:omegabound}) is clearly stronger than this 
``safe'' one. 

Besides these number-theoretical insights made possible by the close
correspondence between the microcanonical statistics of an ideal Bose 
gas and the factorisation problem, there also is a conceptual aspect
of our work: For numbers~(\ref{DPF}) with different, {\em i.e.\/},
distinguishable prime factors, the probability distribution~(\ref{DIS}),
when normalised to unit variance, approaches a Gaussian for $N \to \infty$.
In contrast, if the prime factors are taken to be equal, {\em i.e.\/},
indistinguishable, the distribution remains distinctly non-Gaussian even 
in the asymptotic limit. Thus, we encounter here a fairly nontrivial model
for the occurrence of non-Gaussian statistics.

\appendix

\section{\label{ap:log}Eigenvalues for a logarithmic potential}  

Let us consider the motion of a particle with mass~$m$ in a one-dimensional 
potential
\begin{equation}
   V(x) = V_0 \ln\!\left(\frac{|x|}{L}\right) \; ,
\end{equation}
where $V_0$ and $L$ are positive constants with the dimension of energy 
and length, respectively. Within the quasi-classical Bohr-Sommerfeld 
approximation~\cite{LL}, the quantum mechanical energy eigenvalues
$\varepsilon_\nu$ pertaining to this potential are obtained by setting 
the classical action
\begin{equation}
   I = \frac{1}{2\pi} \oint p \, \rd x 
\end{equation}
equal to $\hbar(\nu + 1/2)$, where $p = p(x)$ denotes the classical
momentum at the position $x$, and $\nu = 0,1,2,\ldots$ is an integer.

Introducing the right turning point $x_\nu$ corresponding to classical motion
with energy $\varepsilon_\nu$,
\begin{equation}
   x_\nu = L \exp\!\left(\frac{\varepsilon_\nu}{V_0}\right) \; ,
\label{REL}   
\end{equation}     
and accounting for the symmetry of the potential, the quantisation 
condition becomes
\begin{eqnarray}
   \frac{\hbar\pi}{2}\left(\nu + \frac{1}{2}\right) & = & 
   \int_0^{x_\nu} \! \rd x \, 
   \sqrt{2m\left(\varepsilon_\nu - V_0\ln\frac{x}{L}\right)}
\nonumber \\ & = &
   \sqrt{2mV_0} \int_0^{x_\nu} \! \rd x \, \sqrt{-\ln\frac{x}{x_\nu}} 
\nonumber \\ & = & 
   \sqrt{2mV_0} \, x_\nu \, \Gamma\!\left(\frac{3}{2}\right) \; ,
\end{eqnarray}
giving
\begin{equation}
   x_\nu = \sqrt{\frac{\pi}{2mV_0}} \hbar \left(\nu + \frac{1}{2} \right) \; .
\end{equation}
Utilising the relation~(\ref{REL}) for expressing the turning point through 
the energy, this latter equation yields the desired approximate eigenvalues
\begin{equation}
   \frac{\varepsilon_\nu}{V_0} = \ln(2\nu + 1) + 
   \ln\!\left(\frac{\hbar}{2L}\sqrt{\frac{\pi}{2mV_0}}\right) \; .
\end{equation}

\section{\label{ap:rec}Derivation of the recursion relation}

The derivation of the recursion relation~(\ref{eq:recpro}) presented here 
is not restricted to logarithmic energy levels and thus generalises the 
derivation previously given in ref.~\cite{Weiss97}.

As in section~\ref{sec:rec}, let~$\Gamma_k({N})$ be the number of possible 
factorisations of~$N$ into $k$ or less natural numbers larger than~$1$. If 
there are less than $k$~factors, say $k-m$, we multiply the product by $1^m$. 
(This emphasises the correspondence with the Bose condensate: The factor 
$1^m$ represents $m$ particles which reside in the ground state, carrying
no energy.) For example, $\Gamma_4(12)$ gives rise to the four factorisations
\begin{eqnarray}
  12 & = & 2  \cdot 2 \cdot 3 \cdot 1   \nonumber \\
     & = & 2  \cdot 6 \cdot 1 \cdot 1   \nonumber \\
     & = & 3  \cdot 4 \cdot 1 \cdot 1   \nonumber \\
     & = & 12 \cdot 1 \cdot 1 \cdot 1   \; .
\label{eq:fact12an}             
\end{eqnarray}
Keeping both $N$ and $k$ fixed, and randomly selecting one of the possible 
factorisations, the probability for the factor~$\nu$ to occur at least
$n$~times is given by
\begin{equation}
    P_{\nu}^{\ge}(n) =
   \left\{\begin{array}{c@{\quad:\quad}l}
   \frac{\Gamma_{k-n}({N}/\nu^n)}{\Gamma_{k}({N})} 
     & {N} \bmod \nu^n =0 \\
   0 & \mbox{else}
   \end{array}
   \right.\; .
\end{equation}
The probability to find the factor~$\nu$ exactly~$n$ times is then 
obtained as a difference,
\begin{equation}
   P_{\nu}(n) = 
   P_{\nu}^{\ge}(n) - P_{\nu}^{\ge}(n+1) \; .
\end{equation}
Next, let $\#_{\nu}$ be the number of occurrences of the factor $\nu$
in some product. Taking the average over all possible products, we obtain
\begin{eqnarray}
   \overline{\#_{\nu}} & \equiv & \sum_{n=1}^{k} n \, P_{\nu}(n) 
\nonumber \\
   & = & \sum_{n=1}^{k} P_{\nu}^{\ge}(n)
\nonumber \\
   & = & \frac1{\Gamma_k({N})}
   \sum_{n=1 \atop N \bmod {\nu}^n = 0}^k
   \Gamma_{k-n}({N}/\nu^n) \; .
\end{eqnarray}
Since there is no possibility to factorise $N > 1$ such that zero factors
are larger than~$1$, we have $\Gamma_0(N>1) = 0$; since, however, there
trivially is such a possibility for $N = 1$, it follows that 
$\Gamma_0(1) = 1$.

By definition of $\#_{\nu}$ we have for every factorisation 
\begin{equation}
   k = \sum_{\nu=1}^{{N}}\#_{\nu} \; ,
\end{equation}
which also holds upon averaging 
($ k = \sum_{\nu=1}^{{N}} \overline{\#_{\nu}} \; $). Hence,
\begin{equation}
   k = \sum_{\nu=1}^{{N}}
   \frac{1}{\Gamma_k({N})}
   \sum_{n=1 \atop N \bmod {\nu}^n = 0}^k
   \Gamma_{k-n}({N}/\nu^n) \; ,
\end{equation}
leading immediately to
\begin{equation}
  \Gamma_k({N}) = \frac{1}{k}\sum_{\nu=1}^{{N}}
  \sum_{n=1 \atop N \bmod {\nu}^n = 0}^k
  \Gamma_{k-n}({N}/\nu^n) \; .
\label{eq:recproan}
\end{equation}
Since the sums are finite, we can safely exchange the order of summation 
and arrive at the recursion relation~(\ref{eq:recpro}).

\section{\label{ap:an}Asymptotics for products of distinct primes}

With an obvious simplification of notation, the recursion 
relation~(\ref{eq:recdp}) for the number of ways to decompose a product 
of~$m$ distinct primes into exactly $k$~integer factors takes the form
\begin{equation}
  \Phi_{m,\,k}  =  k \Phi_{m-1,\,k}+ \Phi_{m-1,\,k-1} \; ;
  \quad \Phi_{m,\,1} = 1 \; , \quad \Phi_{m,\,k>m} = 0 \; .
\end{equation}
We will first show by induction over~$k$ that the solution to this relation 
is given by
\begin{eqnarray}
\label{eq:steffenana}
   \Phi_{m,\,k} = \frac{(-1)^k}{k!}\sum_{\ell=1}^{k}{(-1)^{\ell}
   {k \choose \ell}{\ell}^m} \; .
\end{eqnarray}
For $k = 1$, equation~(\ref{eq:steffenana}) obviously is correct.
Moreover, we have
\begin{eqnarray}
   & & k \Phi_{m-1,\,k} + \Phi_{m-1,\,k-1} 
\nonumber \\ & = & 
   k\frac{(-1)^k}{k!}
   \sum_{\ell=1}^{k}{(-1)^{\ell}{k \choose \ell}{\ell}^{m-1}}
 + \frac{(-1)^{k-1}}{(k-1)!}
   \sum_{\ell=1}^{k-1}{(-1)^{\ell}{k-1 \choose \ell}{\ell}^{m-1}}
\nonumber \\ & = & 
   \frac{(-1)^k}{k!}\left\{
   \sum_{\ell=1}^{k}(-1)^{\ell}k {k \choose \ell}{\ell}^{m-1}
 - \sum_{\ell=1}^{k-1}(-1)^{\ell}k {k-1 \choose \ell}{\ell}^{m-1}
   \right\} \; .
\end{eqnarray}
Using
\begin{equation}
   {k-1 \choose \ell} = \frac{k-\ell}{k}{k \choose \ell} \; ,
\end{equation}
this yields
\begin{eqnarray}
   & & k \Phi_{m-1,\,k} + \Phi_{m-1,\,k-1}
\nonumber\\ & = & 
   \frac{(-1)^k}{k!}\left\{
   \sum_{\ell=1}^{k}(-1)^{\ell}k {k \choose \ell}{\ell}^{m-1}
  -\sum_{\ell=1}^{k-1}(-1)^{\ell}(k-\ell) {k \choose \ell}{\ell}^{m-1}\right\}
\nonumber\\ & = & 
   \frac{(-1)^k}{k!}\left\{ (-1)^{k}k {k \choose k}{k}^{m-1}
  +\sum_{\ell=1}^{k-1}(-1)^{\ell}\ell {k \choose \ell}{\ell}^{m-1}\right\}
\nonumber\\ & = & 
   \frac{(-1)^k}{k!}\sum_{\ell=1}^{k}{(-1)^{\ell} {k\choose \ell}{\ell}^m} \; ,
\end{eqnarray}
which proves the assertion~(\ref{eq:steffenana}).

This result can now be employed to estimate the total number
$\Omega\left(\prod_{i=1}^m p_i \right)$ of factorisations for 
numbers~(\ref{DPF}) containing no identical prime factors:
\begin{eqnarray}
   \Omega\left( \prod_{i=1}^m p_i \right) & = &
   \sum_{k=1}^m \Phi_{m,k}
\nonumber \\ & = &
   \sum_{k=1}^{m} \sum_{\ell=1}^k \frac{(-1)^{k-\ell}}{(k-\ell)!}
   \frac{\ell^m}{\ell!}
\nonumber \\ & = &  
   \sum_{\ell=1}^m \left( \sum_{\nu=0}^{m-\ell} \frac{(-1)^\nu}{\nu!} \right)
   \frac{\ell^m}{\ell!}
\nonumber \\ & = & 
   \sum_{\ell=1}^m \frac{\Gamma(m-\ell+1,-1)}{e\,(m-\ell)!} 
                   \frac{\ell^m}{\ell!} \; , 
\label{eq:frage}
\end{eqnarray}
where $\Gamma(a,b)$ denotes the incomplete Gamma function. The last 
equality here is proven by induction, using the recursive definition
of~$\Gamma(m,-1)$: 
\begin{equation}
   \Gamma(m+1,-1) =(-1)^m e+m\,\Gamma(m,-1) \; ,\qquad
   \Gamma(1,-1)  =  e \; .
\end{equation}

For large $m$, the sum~(\ref{eq:frage}) is dominated by terms with 
$m-{\ell}+1 \gg 1$, so that we may use the asymptotic relation 
$\Gamma(m-\ell+1,-1) = \int_{-1}^\infty {t^{m-\ell} e^{-t}\,{\rm d} t} 
\sim \Gamma(m-\ell+1)$. This leads to the asymptotic equality
\begin{eqnarray}
   \Omega\left(\prod_{i=1}^m p_i \right) & \sim &
   \frac{1}{e} \sum_{\ell=1}^m \frac{\ell^m}{\ell!}
\nonumber \\ & \sim &
   \frac{1}{e} \int_{1}^{m+1} \! \rd x \, \exp[f(x)] \; , 
\label{SAD}
\end{eqnarray}
where, according to Stirling's formula,
\begin{equation}
   f(x) \sim m\ln x - x\ln x + x \; .
\end{equation}   
The saddle-point approximation to this latter integral~(\ref{SAD}) 
then gives
\begin{equation}
   \Omega\left(\prod_{i=1}^m p_i \right) \sim
   \sqrt{2\pi} \exp\!\left[\left(m+\frac{1}{2}\right)\ln m - m
                           + \frac{m}{\ln m} - \ln(\ln m) - 1 \right] ,  
\end{equation}
resulting in the asymptotic bounds
\begin{equation}
   m\ln m - m
   \; < \;
   \ln\left[ \Omega\left(\prod_{i=1}^m p_i \right) \right] 
   \; < \;
   m \ln m \; .
\label{EST}
\end{equation}

\end{document}